\documentclass{ws-procs975x65}

\begin{document}

\title{Directional detection of galactic Dark Matter}

\author{F. MAYET$^*$, J. BILLARD, G. BERNARD,  
G. BOSSON, O. BOURRION, \\ 
C. GRIGNON, O. GUILLAUDIN, C. KOUMEIR, J.~P. RICHER and D. SANTOS}

\address{Laboratoire de Physique Subatomique et de Cosmologie,\\ 
Universit\'e Joseph Fourier Grenoble 1,\\
  CNRS/IN2P3, Institut Polytechnique de Grenoble,\\ 
  53 rue des Martyrs, 38026 Grenoble, France\\
$^*$E-mail: Frederic.Mayet@lpsc.in2p3.fr}

\author{P. COLAS, E. FERER and  I. GIOMATARIS}
 \address{IRFU/DSM/CEA, CE Saclay, 91191 Gif-sur-Yvette cedex, France}

\author{A. ALLAOUA and L. LEBRETON}
 \address{Laboratory for neutron Metrology and Neutron dosimetry,\\ 
 Institute for Radiological Protection and Nuclear Safety,\\ 
 13115 Saint Paul Lez Durance, France}

\begin{abstract}
Directional detection of galactic Dark Matter is a promising  search strategy for discriminating
geniune WIMP events from  background ones. We present technical progress on gaseous detectors as well as recent phenomenological studies, allowing 
the design and construction of competitive experiments.
\end{abstract}


\bodymatter

\bigskip

Direct detection of Dark Matter is entering a new era as several detectors are starting to exclude  the expected supersymetric parameter space \cite{cdms,idm} and new projects of detector are 
planning to scale-up to ton-scale\cite{idm}. 
An alternative strategy to massive detectors, aiming at high background rejection, is the development of detectors providing an unambiguous positive WIMP signal.
This can be achieved by searching for a correlation of the WIMP signal either with the motion of the Earth 
around the Sun, observed as an annual modulation \cite{freese}, or with the direction of the solar motion around the galactic center, 
observed as a direction dependence of the incoming WIMP flux \cite{spergel}, towards ($\ell_\odot = 90^\circ,  b_\odot =  0^\circ$), which
happens to be roughly in the direction of the Cygnus constellation. 
The latter strategy is generally referred to as directional  detection of Dark Matter and several projects of detector are being developed for 
this goal \cite{Drift,MIMAC,MIMAC2,mit,newage,white}. 
Phenomenological studies have shown that not   
only the forward/backward asymetry  can be used to discriminate Dark Matter and background. By studying the 
recoil spatial distribution with various statistical methods \cite{krauss,morgan,green}, useful information may be extracted from such measurements.\\
Recently, a statistical tool, using a map-based likelihood analysis, has been developed \cite{billard} to 
extract information from  data samples of forthcoming directional detectors, thus  changing the search strategy from 
background rejection to  Dark Matter identification method.\\

\begin{figure}[t]
\includegraphics[scale=0.23,angle=90]{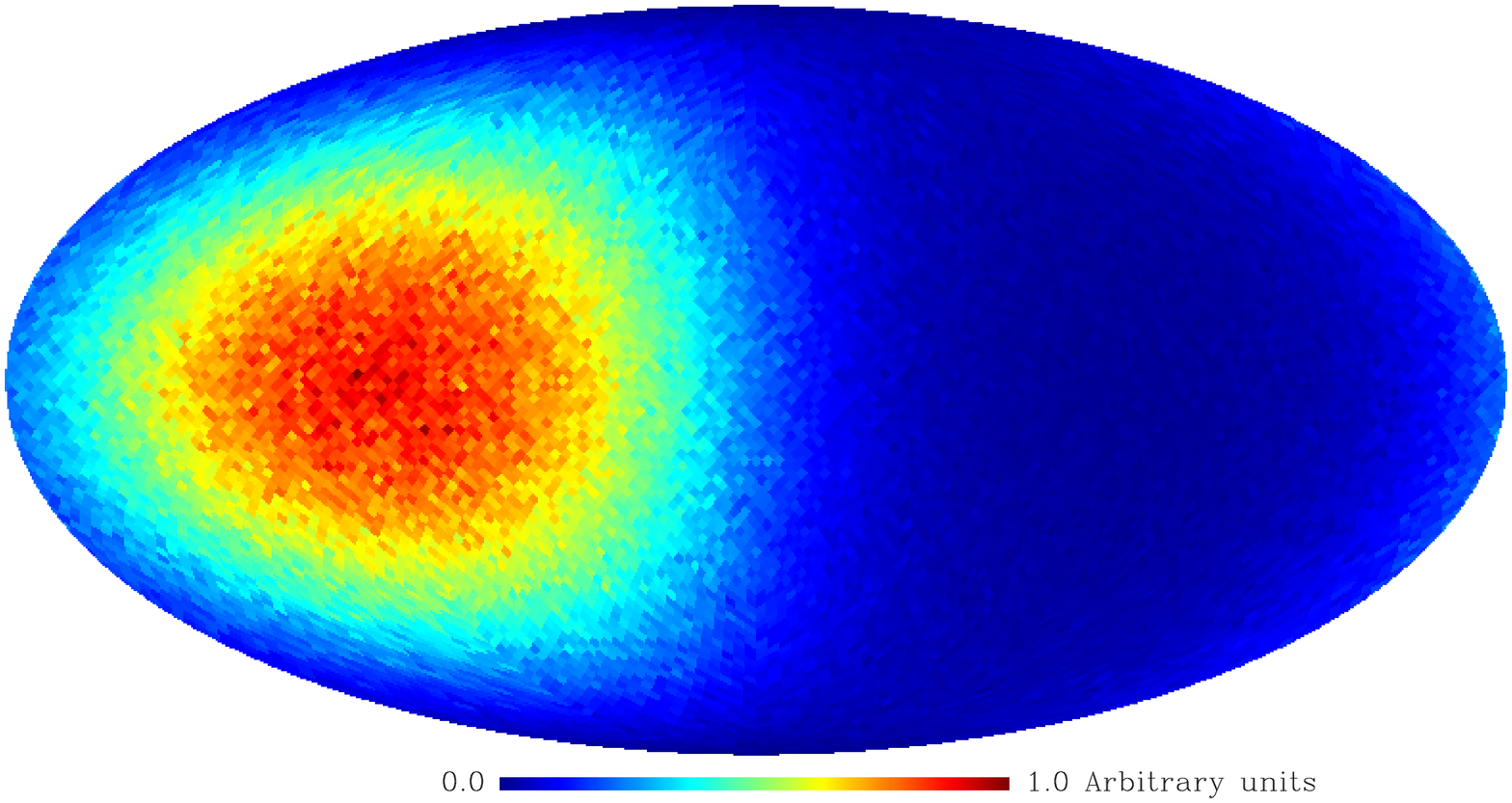}
\includegraphics[scale=0.23,angle=90]{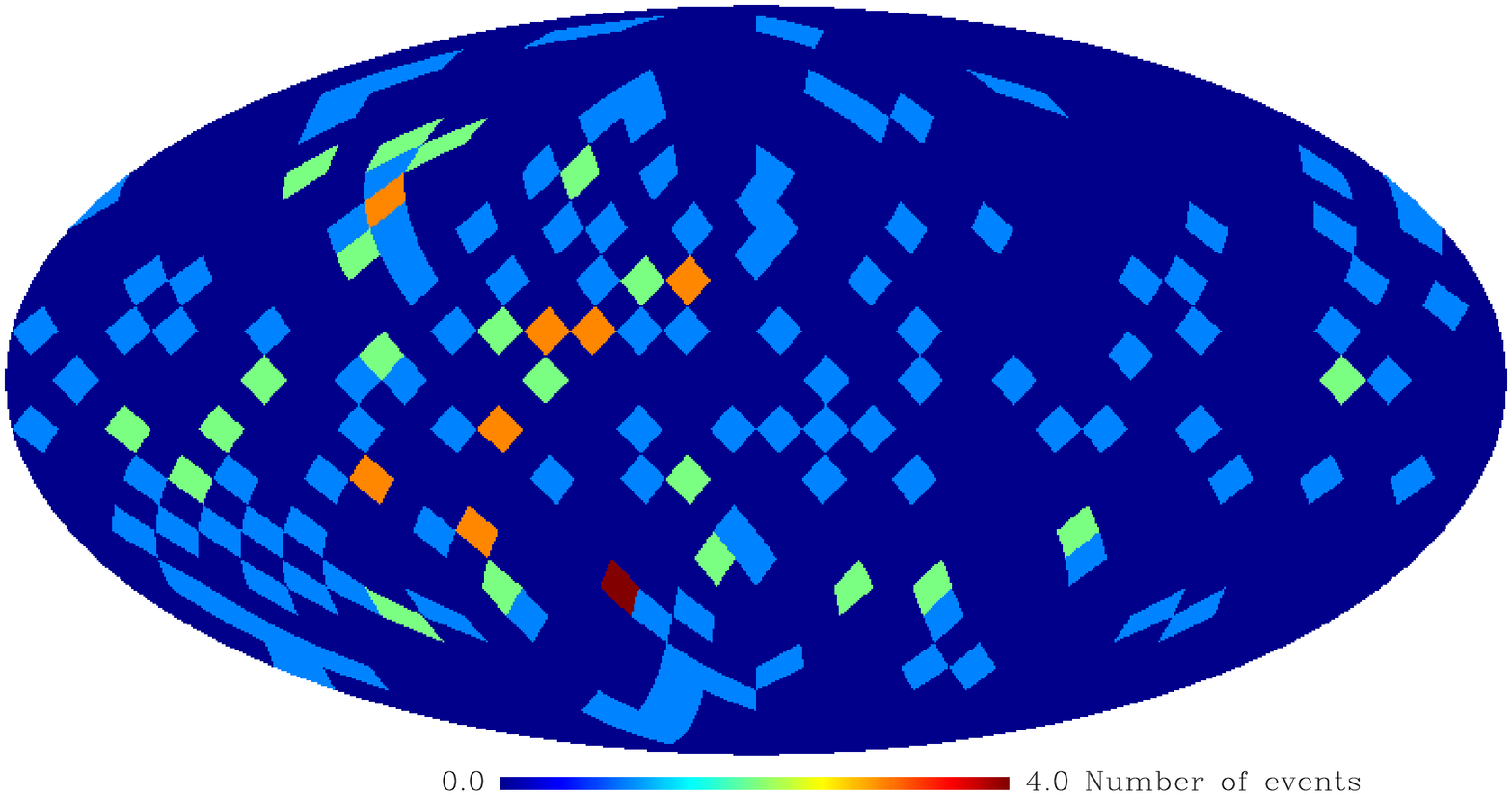}
\caption{Left panel :   WIMP-induced recoil distribution within the framework of an  
 isotropic isothermal spherical Milky Way halo. Right panel :  a typical simulated measurement, {\it i.e.}  100 WIMP-induced recoils and 100 background
events with a low angular resolution. Maps are Mollweide equal area projections. Figure from \cite{billard}.}  
\label{fig:DistribRecul}
\end{figure}

Figure \ref{fig:DistribRecul} presents on the right panel a typical simulated recoil distribution observed by a 
directional detector : $100$ WIMP-induced events and 
$100$ background events. 
For an elastic axial cross-section on nucleon $\rm \sigma_{n} = 1.5 \times 10^{-3} \ pb$ and a $\rm 100 \ GeV.c^{-2}$ WIMP, this corresponds to 
an exposure of $\rm \sim 7\times 10^3  \ kg.day$ in  $\rm ^{3}He$ and $\rm \sim 1.6 \times 10^3 \ kg.day$  in CF$_4$, on their equivalent energy ranges as discussed in \cite{billard}. 
Low resolution maps are used ($N_{\text{bins}} = 768$) to account for low  angular resolution. 
In order to extract information from such a measurement,  a blind likelihood analysis has been developed \cite{billard}, 
allowing to retrieve without any prior,  both the main direction of the incoming events ($\ell, b$) and 
the number of WIMP events contained in the map. It can be  concluded that 
this simulated map contains a signal pointing towards the Cygnus constellation within 10$^\circ$, with $N_{\rm wimp}=96 \pm 15 \ (68 \% \text{CL})$,  
corresponding to a high significance galactic Dark Matter detection. Such a  result could be obtained, with a background rate of 
$\sim 0.07$ kg$^{-1}$day$^{-1}$ and a  10 kg $\rm CF_4$ detector operated  
during  $\sim 5$ months, noticing that the detector should allow 3D 
track reconstruction, with sense recognition  \cite{headtail1,headtail2} down to 5 keV.\\

Several directional detectors are being developed and/or operated : DM-TPC\cite{mit}, NEWAGE\cite{newage}, DRIFT\cite{Drift},
MIMAC\cite{MIMAC}. A detailed overview of the status of experimental efforts devoted to directional
dark matter detection is presented in \cite{white}. Directional detection of Dark Matter requires track reconstruction  of recoiling nuclei down 
to a few keV. This can be achieved with low pressure gaseous detectors \cite{sciolla} and several gases have been suggested : 
$\rm  CF_4,^{3}He+C_4H_{10}$ or  $\rm CS_2$. For these targets, although their detection characteristics may be different 
(e.g. track length, drift velocity and straggling), their directional signature can be  considered as equivalent \cite{billard}.
Both the energy and the track of the recoiling nucleus need to be measured precisely. Ideally, recoiling tracks should be 3D 
reconstructed as the required exposure  is decreased by an order of magnitude between  2D read-out and 3D read-out \cite{green}. 
Sense recognition of the recoil track is also a key issue for
directional detection \cite{headtail1,headtail2}.\\

The MIMAC project is based on a matrix of gaseous $\mu$TPC, filled either with one or several gases amongst 
$\rm ^{3}He$, $\rm CF_4$, $\rm CH_4$ and 
$\rm C_4H_{10}$. Using both $\rm ^{3}He$ and $\rm CF_4$ in a patchy matrix of $\mu$TPC opens the possibility to 
compare the rates for two atomic masses, and to study neutralino interaction separately on neutron and proton. 
With low mass targets, the challenge is   to measure low energy recoils, e.g. below 
$\rm \mathcal{O}  (1-10) \ keV$ for Helium, by means of ionization measurements.  
Accurate energy measurement has been achieved with Helium $\mu$TPC, implying precise knowledge of 
the ionization quenching factor down to sub-keV energies \cite{quenching}. 
3D reconstruction of $mm$ tracks  is also an experimental challenge.
A $\mu$TPC prototype with a 16.5~cm drift space has been developed and successfully operated in surface laboratory and tested 
with mono-energetic neutron fields at the AMANDE facility (IRSN Cadarache) \cite{amokrane}. We use a bulk micromegas \cite{Giomataris}  with a $\rm 3\times 3 \ cm^2$ active area, 
segmented in $300 \ \mu m$ pixels, and a 325~LPI (Line Per Inch) 
weaved 25~$\mu$m thick stainless steel micro-mesh. The $\mu$TPC prototype is equipped with a  
dedicated front end ASIC (BiCMOS-SiGe 0.35 $\mu$m), self-triggered and able to perform a 40 MHz anode 
sampling \cite{Richer:2009pi}. The detection strategy is the following : primary electrons are 
produced by the recoil nucleus and drifted towards the grid in the drift space and then  collected on the pixelized anode. This 
allows to access information on X and Y coordinates.
The Z coordinate is then obtained with a 40~MHz sampling of the anode, providing  the  
electron drift velocity is known. Next step of the MIMAC project is the construction of a small $\mu$TPC matrix to be operated in
underground laboratory.

\end{document}